\documentclass[12pt, journal, comsoc, one column, draftclsnofoot]{IEEEtran}

\usepackage{subcaption}
\usepackage[font=footnotesize]{caption}
\usepackage{amsfonts}
\usepackage{amsthm}
\usepackage{amsmath}
\usepackage{multirow}
\usepackage{array}
\usepackage{bbm}
\usepackage[numbers,sort&compress]{natbib} 
\usepackage{dsfont}
\usepackage{epsfig}
\usepackage{float}
\usepackage[T1]{fontenc}
\usepackage{graphicx}
\usepackage{multicol}
\usepackage{flafter}
\usepackage[linesnumbered,ruled]{algorithm2e}
\usepackage{array}
\newcolumntype{L}[1]{>{\raggedright\let\newline\\\arraybackslash\hspace{0pt}}m{#1}}
\newcolumntype{C}[1]{>{\centering\let\newline\\\arraybackslash\hspace{0pt}}m{#1}}
\newcolumntype{R}[1]{>{\raggedleft\let\newline\\\arraybackslash\hspace{0pt}}m{#1}}

\usepackage{url}

\newtheorem{theorem}{Definition}

\begin{document}


\title{Survey Propagation: A Resource Allocation Solution for Large Wireless Networks}

\author{Andrea~Ortiz,~\IEEEmembership{Member,~IEEE,}
        Daniel~Barragan-Yani
\thanks{A. Ortiz is with the Communication Engineering Lab, Technische Universit\"{a}t Darmstadt, e-mail: a.ortiz@nt.tu-darmstadt.de.}
\thanks{D. Barragan-Yani is with the Department of Physics and Materials Science, University of Luxembourg, e-mail: daniel.barragan@uni.lu}
\thanks{This work has been funded by the Deutsche Forschungsgemeinschaft (DFG, German Research Foundation) Projektnummer 210487104 - SFB 1053 MAKI.}}


\maketitle
\vspace{-1cm}
\begin{abstract}
The ever-increasing number of nodes in current and future wireless communication networks brings unprecedented challenges for the allocation of the available communication resources. This is caused by the combinatorial nature of the resource allocation problems, which limits the performance of state-of-the-art techniques when the network size increases.
In this paper, we take a new direction and investigate how methods from statistical physics can be used to address resource allocation problems in large networks. 
To this aim, we propose a novel model of the wireless network based on a type of disordered physical systems called spin glasses. 
We show that resource allocation problems 
have the same structure as the problem of finding specific configurations in spin glasses. 
Based on this parallel, we investigate the use of the Survey Propagation method from statistical physics in the solution of resource allocation problems in wireless networks.
Through numerical simulations we show that the proposed statistical-physics-based resource allocation algorithm 
 is a promising tool for the efficient allocation of communication resources in large wireless communications networks. 
Given a fixed number of resources, we are able to serve a larger number of nodes, compared to state-of-the-art reference schemes, without introducing more interference into the system.
 \end{abstract}

\section{Introduction}
\label{sec:intro}
%
%
%
%
\subsection{Motivation and Challenges}
The density of current wireless communication networks is rapidly increasing. This is due to the fact that the networks are not only formed by human-operated devices, but also by the millions sensors and actuators that form the Internet of Things (IoT) \cite{Bockelmann2018}.
To support such communication networks, one of the key questions to be addressed is how to assign the available communication resources at large scale.
This is a challenging task because allocation problems 
often lead to combinatorial optimization formulations, like graph-coloring, which are known to be NP hard. 
Moreover, the complexity of the allocation problem is not constant. It depends on the number of available resources, the number of devices, and the network topology. 
Consequently, optimal solutions can only be found, within reasonable computation time, when a small number of resources and devices is considered. 
Although approximation and heuristic techniques have been develop to find suboptimal allocation  solutions \cite{Ramanathan1999, Mahonen2005, Wang2005, Wu2020, Zhu2017, Kim2013, Wang2017}, their performance drifts away from the optimal as the network size increases and the number of available resources remains constrained.

%
%
%

The study of networks with thousands, or even millions of devices, is a new research direction in wireless communications.
However, in other study fields it is the standard set-up. Our purpose is to exploit the expertise developed in such cases. 
We focus on statistical physics, which aims at the description of the physical properties of macroscopic systems, i.e. systems consisting of a very large number of particles, like atoms or molecules \cite{Mandl1988}.
In order to handle this type of systems, statistical physicists have developed methods that use the statistical distribution of the microscopic behavior of the particles that form them. By drawing a parallel between wireless networks and physical systems, e.g., by considering that the nodes in a wireless network can be seen as the particles in a physical system, such methods can be adapted to address resource allocation problems.
Concepts from statistical physics, e.g., Gibbs sampling, Markov random fields, and phase transitions, have been successfully used to solve communication related problems \cite{Kauffmann2007, Chen2010, Sem2014, Girnyk2014, Michalopoulou2013}.
However, we follow a different approach and, for the first time, study the application of the Survey Propagation method from statistical physics. The main advantage of this approach is that it considers the interaction between the composing elements of the network. Thus, it enables the solution of assignment problems in large networks when the number of resources is limited.

\subsection{Related Work}
Due to the combinatorial nature of resource allocation problems, finding the optimal allocation strategy requires solving complex integer optimization problems. 
Therefore, to address real communication networks, researchers have follow mainly two paths: (i) reducing the complexity of the algorithms, e.g., by developing heuristic solutions  based on graph coloring \cite{Ramanathan1999, Mahonen2005, Wang2005, Wu2020, Zhu2017, Kim2013, Wang2017}, and more recently, machine learning \cite{Jia2020, Ortiz2019, He2019}, or (ii) by developing clustering techniques to divide the original problem into multiple smaller ones. Thus, enabling the use of optimization techniques to solve each of the smaller problems. \cite{Ling2018, Lin2018, Alcaraz2018, Yemini2019}.

The main advantage of the heuristic solutions in \cite{Ramanathan1999, Mahonen2005, Wang2005, Wu2020, Zhu2017, Kim2013, Wang2017} is their reduced complexity compared to optimization-based techniques. However, their main drawback is that their proposed graph coloring techniques are based on greedy approaches whose performance decreases when the network size increases. 
The machine learning techniques in \cite{Jia2020, Ortiz2019, He2019} face a similar problem. Supervised learning approaches, like \cite{Jia2020}, require the computation of optimal allocation solutions in order to train the artificial neural networks. Consequently, the developed architectures are trained for a small number of devices and resources.
In the case of reinforcement learning \cite{Ortiz2019, He2019}, an increasing number of devices and resources in the network impacts the learning speed of the algorithm because the number of possible solutions is an exponential function of these two parameters.

To overcome the fact that the performance of current heuristic approaches do not scale with the network size, clustering techniques have been considered in \cite{Ling2018, Lin2018, Alcaraz2018, Yemini2019}. The main idea of these works is to reduce the size of the problem by dividing the network into smaller ones. 
Although clustering solutions indeed reduce the number of resources and devices to be considered, the current trend in wireless communications indicates that even the size of these clusters will increase. Hence, the initial challenge of finding resource allocation solutions for large networks still remains open.


\subsection{Contributions}
Our goal is to fill the void regarding resource allocation algorithms for large networks by investigating methods from statistical physics, in particular, methods developed for a type of disordered physical systems called spin glasses. 
We show that resource allocation problems, e.g., time, code or frequency assignment, have the same structure as the problem of finding specific configurations in spin glasses. 
Both are combinatorial optimization problems that can be formulated as graph-coloring problems, and both can be modeled in a general manner as Constraint Satisfiability Problems (CSPs).
As it will become clear in the next sections, modeling the resource allocation problems as CSP facilitates the adaptation of methods from statistical physics to wireless networks.
Specifically, we investigate the use of the Survey Propagation, a statistical physics method proposed in \cite{Mezard2002}, as a novel heuristic approach to solve hard CSPs.
Its underlying concept emerged from the study of the interaction between the composing elements of spin glasses, i.e., the spins.
The central idea of Survey Propagation is to propagate statistical information, called surveys, about the spins in the system in order to iteratively assign values to the variables in the CSP which describe the spin's state. When all the variables in the CSP are assigned, i.e., the state of all spins have been determined, the desired configuration of spins is said to be found.


To illustrate the advantages of Survey Propagation, we consider a scenario formed by a large number of interconnected nodes. These nodes can be, for example, base stations in a cellular network, aggregators in a massive machine-type communication scenario, or IoT devices. Our goal is to minimize the interference between neighboring nodes through the allocation of time-frequency resources when only information about the network topology is available. 
To this aim, we model the wireless network as a spin glass and show that that the resource allocation problem can be written as a CSP. Using this model, we apply Survey Propagation to find the resource allocation solution that minimizes the interference. 
The contributions of the paper can be summarized as follows:
\begin{itemize}
\item We provide the first application of the Survey Propagation method from statistical physics to a wireless communication network. Moreover, we propose a model for wireless communication networks based on the spin glass model.
\item We formulate the resource allocation problem as a CSP and represent it as a factor graph. We show that these formulation and representation facilitate the application of Survey Propagation to find the resource allocation solution that minimizes the interference when only information about the network topology is available. 
\item We propose a resource allocation algorithm based on Survey Propagation which minimizes the interference between neighboring nodes in large networks. Additionally, we discuss the convergence properties of the proposed approach by studying the structure of the considered wireless network.
\item Using numerical simulations, we show that the proposed algorithm outperforms state-of-the-art resource allocation approaches, namely, Belief Propagation and greedy heuristics, as well as low-complexity strategies such as random allocation.
\end{itemize}

%
The rest of the paper is organized as follows. In Section \ref{sec:spinGlass}, we discuss parallels between spin glasses and wireless communication networks. Specifically, we give a brief introduction to spin glasses and show how the problem of finding minimal energy configurations in spin glasses is analogous to finding resource allocation solutions that minimize the interference in wireless networks. In Section \ref{sec:probFormulation}, the considered system model is described and the resource allocation problem is formulated as a CSP. Additionally, practical application examples for the proposed spin glass model of the wireless network are presented.
The use of Survey Propagation to solve the resource allocation problem in large wireless networks and the proposed algorithm are explained in Section \ref{sec:SP}. 
In Section \ref{sec:conv}, the convergence properties of Survey Propagation 
are discussed. Numerical performance results are presented in Section \ref{sec:simResults} and Section \ref{sec:conclusions} concludes the paper. 

\section{Spin Glasses and Wireless Communication Networks}
\label{sec:spinGlass}
\subsection{Spin Glasses}

\begin{figure}[t]
     \centering
         \includegraphics[width=0.28\textwidth]{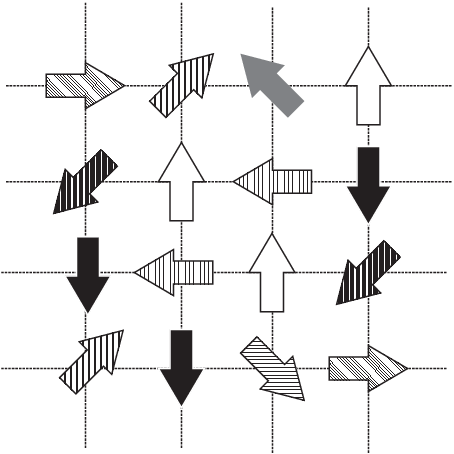}
         \caption{Simplistic depiction of a spin glass in a grid.}
     \label{fig:spinGlass}
\end{figure}

Spin glasses are a fundamental type of disordered systems studied in statistical physics. They refer to magnetic systems in which conventional ferromagnetic or antiferromagnetic long-range orders cannot be established due to some structural disorder. As a result, the magnetic moments or spins composing the system prefer to be arranged in random directions \cite{Binder1986, Mezard1987, Nishimori2001}. One way to imagine, or model, spin glasses is to assume that the spins are located 
at fixed points in a regular lattice, see Fig. \ref{fig:spinGlass}, with disorder introduced via random couplings, or interactions, that follow a suitable distribution \cite{Edwards1975, Sherrington1975, Nishimori2001}.

The spin glass model can very well be applied to describe communication systems, specifically, a large wireless communication network. 
In the spin glass model of the communication network, each spin represents one communication node and its direction corresponds to the orthogonal resource allocated to it, e.g., a frequency band or a time slot. 
For example, in Fig. \ref{fig:spinGlass}, we assume eight orthogonal resources, this means, eight possible directions (North, South, East, West, North-East, North-West, South-East and South-West). To highlight the different directions, we have used different colors and patterns for each of them.
The random coupling between two spins corresponds, in the communication case, to the interference caused by a pair of neighboring transmitting nodes using the same orthogonal resource.
Since nature has an invariable tendency to satisfy the principle of minimum energy \cite{Callen85}, statistical physicists are often interested in finding minimal energy configurations in order to describe the equilibrium properties of relevant systems, like spin glasses \cite{Edwards1975, Sherrington1975, Nishimori2001}. From a communication perspective, we have a similar aim. Our goal is to find the resource allocation solution (orientation of spins) which  minimizes the interference in the system (yields the minimum energy in the spin glass).,

\subsection{Constraint Satisfiability Problem (CSP)}
\label{sec:CSP}
CSPs are at the core of combinatorial optimization theory and deal with the question of whether a set of constraints $\Gamma$ between discrete variables $\mathcal{X}$ can be simultaneously satisfied.
Each constraint is a clause formed by the logical disjunction (OR) of a subset of the variables or their negations \cite{Mezard2002}.
The solution of the CSP is an assignment for the variables that guarantees that all the constraints  in the problem are satisfied.
CSPs are strongly related to the theory of spin glasses because the problem of finding the minimal energy configuration of the spin glass model can be written as a CSP. 
The clauses in the CSP are associated to the interactions between neighboring nodes and the orientation of each spin.
The minimal energy configuration is an assignment that satisfies all clauses. 
Similarly, the resource allocation problem in wireless networks can be formulated as a CSP. Details of such formulation are presented in Section \ref{sec:probFormulation}.

\begin{figure}[ht]
     \centering
         \includegraphics[width=0.7\textwidth]{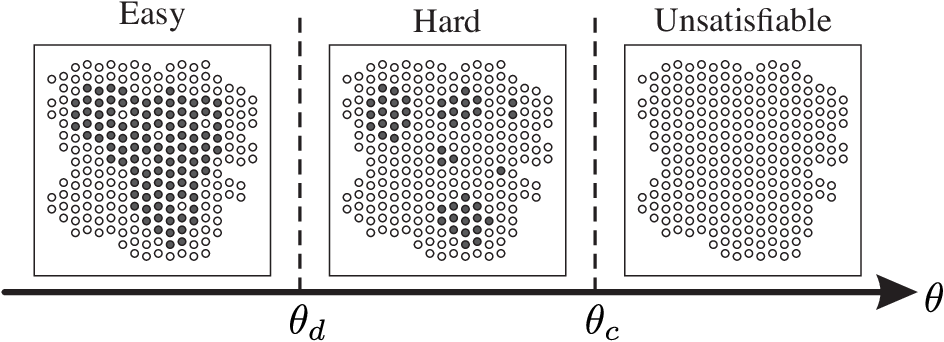}
         \caption{Change in the solution space as $\theta$ increases. Figure based on the work of \cite{Zdeborova2007, Maneva2007}.}
     \label{fig:cluster}
\end{figure}

Finding the minimal energy configuration and, similarly, the resource allocation problem for interference minimization are combinatorial problems. 
The design of algorithms to find configurations that fully satisfy the CSP and the determination of whether a given CSP can be satisfied, is a challenging task \cite{Mezard2002, Mezard2002B, Maneva2007}.
This is because finding a solution heavily depends on how constrained is the particular problem at hand.
Let us define the ratio $\theta$ between the number $|\mathcal{X}|$ of variables and the number $|\Gamma|$ of constraints in the CSP as 
\begin{equation}
\theta = \frac{|{\Gamma}|}{|\mathcal{X}|}. 
\end{equation}
There exists a critical threshold $\theta_c$ for which the CSP becomes unsolvable.
When $\theta < \theta_c$, the CSP can be satisfied. Conversely, when $\theta > \theta_c$ the CSP is unsatisfiable \cite{Mezard2002B, Mezard2002, Zdeborova2007}.
Note that within the satisfiable region $\theta < \theta_c$ the complexity of finding a solution is not constant. 
Physicists have found that there exists an intermediate threshold $\theta_d$ that specifies a region $\theta_d<\theta<\theta_c$ where a CSP is still satisfiable but the solution is hard to find \cite{Mezard2002}\footnote{Other works have identified more intermediate thresholds to make a finer characterization of the behavior of the solution space \cite{Zdeborova2007}.}.
Such difficulty comes from the fact that when $\theta$ increases, the solution space becomes clusterized, as depicted in Fig. \ref{fig:cluster}. 

In the figure, each circle represents a solution of the CSP. 
Filled circles mean the allocation satisfies the CSP, and empty circles denote that the allocation does not satisfy all the clauses.
For $\theta<\theta_d$, the satisfying allocation (filled circles) are close to each other forming one single cluster.
However, for larger values of $\theta$, groups of satisfying allocations are apart forming many smaller clusters. The number of clusters increases exponentially until the CSP becomes unsatisfiable \cite{Mezard2002}.
In the region $\theta_d<\theta<\theta_c$, the allocations in separate clusters are far apart and moving from one allocation in a cluster to some other allocation in another cluster requires simultaneously changing the value of many of the considered variables \cite{Braunstein2005, Maneva2007}.

For resource allocation in wireless communication networks, the solution of the CSP is highly dependent on the network topology, the density of nodes and the number of available orthogonal resources.
The threshold $\theta_c$ indicates the minimum number of orthogonal resources needed to find an allocation solution that minimizes the interference.
Moreover, finding solutions in the range $\theta_d \leq \theta \leq \theta_c$ allows us to reduce the number of required orthogonal resources compared to solutions in the range $\theta<\theta_d$.
In this paper,  we exploit the Survey Propagation method to find a solution to the resource allocation problem, especially in the range $\theta_d \leq \theta \leq \theta_c$. 
The application of Survey Propagation to the resource allocation problem is described in Section \ref{sec:SP}.
The parallel between spin glasses and wireless communication networks is summarized in Table \ref{table:parallel}, where the main components of a spin glass model are mapped in a one-to-one manner to the main elements in a wireless network. 

\begin{table}[ht]
	 \centering{\scriptsize{
	\begin{tabular}{>{\centering\arraybackslash}m{2cm} >{\centering\arraybackslash}m{4cm}  >{\centering\arraybackslash}m{4cm} }
	\hline \hline
	& Spin Glass & Wireless Network \\
	\hline \hline
	& Spin& Communication node \\
	Scenario & Orientation   & Communication resource\\	
	& Interaction  & Interference\\	
	\hline
	Goal & Find the minimal energy configuration  & Find the allocation that minimizes the interference\\
	\hline
	Problem Formulation & \multicolumn{2}{c}{Constraint Satisfiability Problem}\\
	\hline
	Solution Method & \multicolumn{2}{c}{Survey Propagation}\\
	\hline
	\end{tabular}}}
		\caption{Parallel between spin glasses and wireless networks.}
	\label{table:parallel} 
\end{table}

%


%
%
%
%
%
%

\section{Resource Allocation as a Constraint Satisfiability Problem}
\label{sec:probFormulation}

\subsection{System Model}
\label{sec:systemModel}
In this section, the considered system model is introduced. Additionally, we provide practical application scenarios of our model, introduce CSPs, and describe its corresponding factor graph representation. To facilitate the readability of the paper, a table with the summary of the notation is provided in Table \ref{tab:notations}.

We consider a multi-cell network formed by a large number $I \in \mathbb{N}$ 
of base stations, each of them serving a group of users.
Without loss of generality, we assume that $N$ time-frequency resources are divided into $Q \in \mathbb{N}$ resource pools, each of them containing $N/Q$ time-frequency resources.
Every base station $i = 1,...,I$ is assigned a resource $q=1,...,Q$. The time-frequency resources within the assigned resource pool $q$ are used by base station $i$ to provide communication to the end users associated to it. 
To minimize the interference in the network, we aim at exclusive resource allocation, i.e., neighboring base stations cannot share the same resource pool $q$. 
Two base stations $i$ and $j$ are said to be neighbors if the received power $p^{\mathrm{Rx}}_{i,j} \in \mathbb{R}$ of a test interference signal $s_{i,j} \in \mathbb{C}$ sent from $i$ and received at $j$ is above a given threshold $\mu \in \mathbb{R}$.
Consider the channel coefficient $h_{i,j} \in \mathbb{C}$ for the link between base stations $i$ and $j$ and assume that each base station uses a fixed transmit power $p^{\mathrm{Tx}} \in \mathbb{R}$.
The power $p^{\mathrm{Rx}}_{i,j}$ of the received interference signal is calculated as
\begin{equation}
p^{\mathrm{Rx}}_{i,j} = |h_{i,j}|^2 p^{\mathrm{Tx}} + \sigma^2,
\end{equation}
where $\sigma^2$ is the noise power at each base station.
Additionally, channel reciprocity between base stations is assumed, i.e., $h_{i,j}=h_{j,i}$. Consequently, $p^\mathrm{Rx}_{i,j} = p^\mathrm{Rx}_{j,i}$.

We model the network described above as a graph $G=(\mathcal{V}, \mathcal{E})$, where $\mathcal{V}$ is the set of vertices and $\mathcal{E}$ is the set of edges. 
In our case, the set $\mathcal{V}$ of vertices contains all the base stations $i$ in the network, such that $\mathcal{V}=\{1,...,i,...,I\}$ and $|\mathcal{V}|=I$. The set $\mathcal{E}$ of edges $e=(i,j)$ contains the links between neighboring base stations $i,j$, i.e., $\mathcal{E}=\{e = (i,j) : i,j\in \mathcal{V}, p^\mathrm{Rx}_{i,j} \geq \mu\}$.

\begin{table}[H]
\footnotesize
\centering
\begin{tabular}{@{}cp{6,6cm} | cp{6,6cm}@{}}
\hline 
\hline
\renewcommand{\arraystretch}{2}
{Symbol} & {Description} & {Symbol} & {Description}\\
\hline
\hline
$\mathcal{A}$ & Set of $\alpha_i$ clauses & $W^+_{x_{i,q}}$ & Bias of $x_{i,q}$ towards "0" \\
$\mathcal{B}$ & Set of $\beta_e$ clauses & $W^+_{x_{i,q}}$ & Bias of $x_{i,q}$ towards "1" \\
$C$ & Cost function of the CSP & $x_{i,q}$ & Binary variable in the CSP, variable node in factor graph \\
$\mathbb{C}$ & Set of complex numbers& $\mathbf{X}$ & Resource allocation solution \\
$e$ & Index for edges $(i,j)$ & $\mathcal{X}$ & Set of all variables $x_{i,q}$ \\
$\mathcal{E}$ & Set of all edges& $\mathcal{X}^*$ & Set of variables $x_{i,q}$ whose value has been fixed\\
$G$ & Graph & $\mathcal{X}(\alpha_i)$ & Set of all $x_{i,q}$ considered in clause $\alpha_i$ \\
$h_{i,j}$ & Channel coefficient between base stations $i$ and $j$ & $\alpha_i$ & Clause ensuring base station $i$ is allocated a resource pool \\
$i$ & Index for the base stations& $\beta_e$ & Clause to avoid sharing the same resource pool \\
$I$ & Total number of base stations & $\gamma$ & Constraint satisfaction problem\\
$j$ & Auxiliary index for the base stations & $\Gamma$ & Set of all clauses, or of all the functional nodes\\
$\mathbb{N}$ & Set of natural numbers & $\Gamma(x_{i,q})$ & Set of functional nodes connected to $x_{i,q}$\\
$p_{i,j}^\mathrm{Rx}$ & Power of the received interference signal from $i$ to $j$ &  $\Gamma^\mathrm{s}_{\alpha_i}(x_{i,q}) $ & Set of functional nodes that tend to make $x_{i,q}$ satisfy $\alpha_i$  \\
$p^\mathrm{Tx}$ & Transmit power &  $\Gamma^\mathrm{u}_{\alpha_i}(x_{i,q}) $ & Functional nodes that tend to make $x_{i,q}$ not satisfy $\alpha_i$ \\
$q$ & Index for the resource pools &  $\delta$ & Hyperbolicity of a graph \\
$Q$ & Total number of resource pools & $\epsilon$ & Convergence threshold for Survey Propagation\\
$\mathbb{R}$ & Set of real numbers &$\eta_{\alpha_i\rightarrow x_{i,q}}$ & Survey sent from functional node $\alpha_i$ to variables node $x_{i,q}$ \\
$s_{i,j}$ & Interference signal from $i$ to $j$ &  $\lambda$ & Density of the point poisson process\\ 
$t$ & Index for the iteration of the algorithm & $\pi_{x_{j,q}\rightarrow \alpha_i}$ & Survey sent from $x_{i,q}$ to functional node $x_{i,q}$ \\
$t_\mathrm{SP}$ & Index for the iteration of Survey Propagation &  $\mu$ & Power threshold to define neighborhoods\\
$T$ & Maximum number of iterations of the proposed algorithm & $\sigma^2$ & Received noise power \\
$T_\mathrm{SP}$ & Maximum number of iterations of Survey Propagation &  $\theta$ & Ration between clauses and variables\\
$\mathcal{V}$ & Set of vertices &&\\
\hline
\end{tabular}
\caption{Table of notations}
\label{tab:notations}
\end{table}

\subsection{Examples of Application Scenarios}
The scenario described in Section \ref{sec:systemModel} is written in a generic way to cover the different application scenarios that can be addressed using the spin glass model of the wireless network and the resource allocation method proposed in this paper. In the following, we consider two particular examples and show how the system model can be adapted for such cases.
\subsubsection{Massive Machine-Type Communications}
In massive machine type communications, a very large number of interconnected devices is considered. The task is to assign the available communication resources tot enable the communication between the devices and the serving bae station, and minimize the interference, In this case, the set $\mathcal{V}$ of communication nodes will be formed by the $I$ machine-type devices, and the set $\mathcal{E}$ of edges includes the links between neighboring interfering devices.
Note that even if clustering techniques, like data aggregation, are considered, the proposed system model still can be applied to the massive machine type scenario. In this case, the resource allocation is done considering the aggregators are the ones forming the set $\mathcal{V}$ of communication nodes. 
\subsubsection{Pilot Allocation in Cell-Free Massive Multiple-Input Multiple-Output (MIMO)}
Cell-free massive MIMO consists on a large number of distributed access points jointly providing access to a small number of users using the same communication resources \cite{Liu2020}. In order to achieve this, channel estimation is needed in uplink and downlink. However, due to the large number of access points, ensuring the orthogonality of the pilot sequences required for channel estimation is a challenging task. By considering that the access points form the set $\mathcal{V}$ of communication nodes, and that the $Q$ resources correspond to the orthogonal pilot sequences, the proposed system model can be applied for the pilot allocation problem. Furthermore, the proposed resource allocation method can be used to minimize the pilot contamination.

\subsection{Constraint Satisfiability Problem}
Our goal is to allocate the available resource pools $q=1,...,Q$ among the base stations in order to minimize the interference between neighboring ones. 
In this section, we show how this problem can be formulated as a CSP. As it will be described in Section \ref{sec:SP}, this formulation allows us to use the Survey Propagation method to solve the resource allocation problem.

Consider the graph $G=(\mathcal{V}, \mathcal{E})$ introduced in Section \ref{sec:systemModel}.  
The corresponding CSP problem, denoted by $\gamma$, is formed by a set $\mathcal{X} = \{ x_{i,q} : i \in  \mathcal{V}, q={1,...,Q} \}$ of binary variables $x_{i,q} \in \{0,1\}$ and a set of constraints $\Gamma$. 
Each variable $x_{i,q}$ indicates whether resource $q$ is allocated to base station $i$ or not. 
The constraints in $\Gamma$ ensure that all base stations are allocated one resource and that neighboring base stations do not share the same resource.
The goal is to find a resource allocation solution $\mathbf{X} = \{(x_{1,1},...,x_{1,Q})^\mathrm{T},...,(x_{I,1},...,x_{I,Q})^\mathrm{T},\}$, where $(\cdot)^\mathrm{T}$ is the transpose operation, that ensures all the constraints in $\Gamma$ are satisfied, in other words, $\gamma = \mathtt{true}$.

In order to translate the resource allocation problem into a CSP, two requirements need to be considered, i.e., each base station has to be assigned only one resource and neighboring base stations cannot share the same resource. 
As a result, two types of constraints are included in the set $\Gamma$.
We denote these types of constraints as $\alpha_i$ and $\beta_e$. The constraints $\alpha_i$ are defined for every base station $i \in \mathcal{V}$ and ensure that one resource is allocated to base station $i$. Each $\alpha_i$ is defined as the logical disjunction of the variables $x_{i,q}$ for $q=1,...,Q$ as
\begin{equation}
\alpha_i = (x_{i,1} \lor x_{i,2} \lor ... \lor  x_{i,Q}).
\label{eq:alpha_i}
\end{equation}
The set containing the constraints $\alpha_i$  is defined as $\mathcal{A}=\{\alpha_i:i\in\mathcal{V}\}$.
The constraints $\beta_e$ guarantee exclusive allocation, i.e., neighboring base stations do not share the same resource. $\beta_e$ is defined for each neighboring pair of base stations $e=(i,j), \; e \in \mathcal{E}$ as the logical conjunction (AND) of the clauses $(\bar{x}_{i,q} \lor \bar{x}_{j,q})$ which ensure each resource $q$ is not shared. Specifically, $\beta_e$ is defined by considering the negation $\bar{x}_{i,q}$ of variables $x_{i,q}$ as
\begin{equation}
\beta_e = (\bar{x}_{i,1} \lor \bar{x}_{j,1}) \land (\bar{x}_{i,2} \lor \bar{x}_{j,2}) \land ... \land  (\bar{x}_{i,Q} \lor \bar{x}_{j,Q}).
\label{eq:beta_e}
\end{equation}
The set containing all the $\beta_e$ constraints is defined as $\mathcal{B}=\{\beta_e:e\in\mathcal{E}\}$.
The CSP problem $\gamma$ is then written as the logical conjunction of all the $\alpha_n$ and $\beta_e$ constraints as 
\begin{align}
\gamma = \left(\bigwedge_{i\in \mathcal{V}}\alpha_i\right) \wedge \left( \bigwedge_{e\in \mathcal{E}}\beta_e \right).
\label{eq:CSP}
\end{align}
In summary, the CSP $\gamma$ is formed by $|\mathcal{X}|=Q|\mathcal{V}|$ variables and $|\Gamma|=|\mathcal{V}| + Q|\mathcal{E}|$ constraints.

For the CSP $\gamma$ defined above, let us define the cost function $C$ as the number of unsatisfied clauses in \eqref{eq:CSP}. A resource allocation $\mathbf{X}$ is optimal when $\gamma(\mathbf{X})=\mathtt{true}$ and consequently, $C=0$. This means, there is no interference between neighboring base stations because resources are not shared among them. 

\subsection{Factor Graph Representation}
\begin{figure}[t]
     \centering
     \begin{subfigure}[b]{0.4\textwidth}
         \centering
         \includegraphics[width=\textwidth]{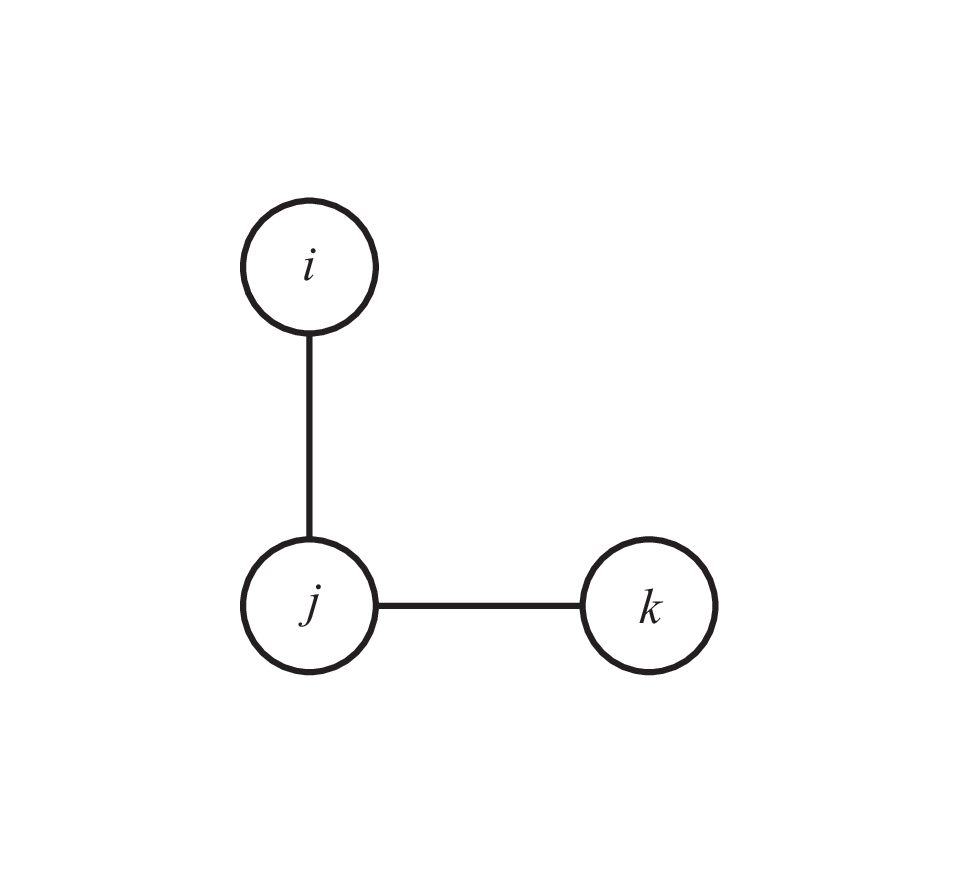}
         \caption{Conventional graph representation.}
         \label{fig:simpleGraph}
     \end{subfigure}
     \hfill
     \begin{subfigure}[b]{0.4\textwidth}
         \centering
         \includegraphics[width=\textwidth]{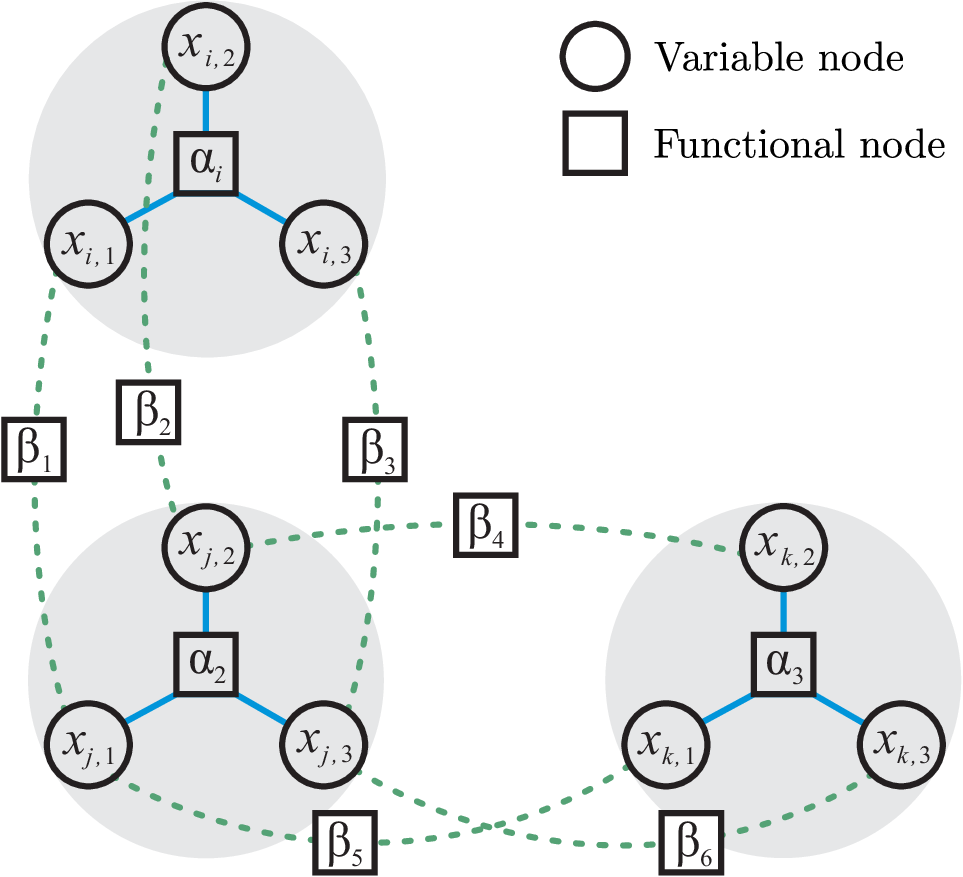}
         \caption{Factor graph representation.}
         \label{fig:extGraph}
     \end{subfigure}
     \caption{Conventional graph and factor graph representations of a network with $I=3$ base stations and $Q=3$ available resources.}
     \label{fig:graphExample}
    \end{figure}

A better understanding of $\gamma$ can be achieved by using the factor graph representation of the CSP \cite{Braunstein2006}. 
A factor graph is a bipartite graph formed by two types of vertices, namely, variable nodes and functional nodes\footnote{Note that the variable nodes and functional nodes do not correspond to the base stations in the wireless network, but to the variables and constraints associated to the CSP.}. 
Considering the CSP, the factor graph representation provides a graphical description of the different constraints and the relation between the base stations. Moreover, it is helpful to explain the Survey Propagation method presented in Section \ref{sec:SP}.
This means, the factor graph does not represent the wireless networks but the constraints that need to be fulfilled in order to achieve minimum interference.

Consider the example in Fig. \ref{fig:graphExample}. On the left side, in Fig. \ref{fig:simpleGraph}, a small network with three base stations is depicted as a graph. The set of vertices is given by $\mathcal{V}=\{i,j,k\}$ and the set of edges by $\mathcal{E}=\{(i,j), (j,k)\}$.
To build the corresponding factor graph, the constraints $\alpha_i \in \mathcal{A}$ and $\beta_e \in \mathcal{B}$, defined in \eqref{eq:alpha_i} and \eqref{eq:beta_e} respectively, are considered.
Specifically, the variable nodes, depicted with circles in Fig. \ref{fig:extGraph}, are the variables $x_{i,q} \in \mathcal{X}$ that determine whether resource $q$ is allocated to base station $i$ or not. 
The set of all variables nodes is then the set $\mathcal{X}$.
The functional nodes, shown with squares in Fig. \ref{fig:extGraph}, correspond to the constraints  $\alpha_i \in \mathcal{A}$ and $\beta_e \in \mathcal{B}$ defined in \eqref{eq:alpha_i} and \eqref{eq:beta_e}, respectively.
The set of all functional nodes is $\Gamma = \mathcal{A} \cup \mathcal{B}$.
An edge between a variable and a functional node means that the variable node is included in the clause represented by the functional node. 
We denote by $\mathcal{X}(\alpha_i)$ the set containing all variable nodes $x_{i,q}$ considered in clause $\alpha_i$\footnote{In case the considered functional node is $\beta_e$ instead of $\alpha_i$, the argument is changed accordingly.}. 
Similarly, we denote by $\Gamma(x_{i,q})$ the set of functional nodes $\alpha_i$ and $\beta_e$ to which the variable node $x_{i,q}$ is connected.
The type of line in the factor graph representation indicates whether the variable $x_{i,q}$ or its negation $\bar{x}_{i,q}$ is used. In Fig. \ref{fig:extGraph}, the edges connecting the variable nodes with the functional nodes $\alpha_i$ are depicted with solid lines because the variables $x_{i,q}$ are considered in this clause. 
On the contrary, the edges connecting the variable nodes with the $\beta_e$ constraints are depicted with dashed lines because the negated variables $\bar{x}_{i,q}$ are considered.
Note that edges between two variable nodes are not possible since variable nodes can only be related through clauses, i.e., through functional nodes. 

Considering a functional node $\alpha_i$ and a variable node $x_{i,q}$, we define $\Gamma^\mathrm{s}_{\alpha_i}(x_{i,q}) $ and $\Gamma^\mathrm{u}_{\alpha_i}(x_{i,q})$ as the sets containing all the functional nodes connected to $x_{i,q}$ which tend to make variable node $x_{i,q}$ satisfy ($\Gamma^\mathrm{s}_{\alpha_i}(x_{i,q}) $) or not satisfy ($\Gamma^\mathrm{u}_{\alpha_i}(x_{i,q}) $) clause $\alpha_i$, respectively.  
From the definition of $\alpha_i$ and $\beta_e$, it is clear that a variable node $x_{i,q}$ can be connected to multiple $\beta_e$ functional nodes but only to one $\alpha_i$ node. Consequently, for functional nodes $\alpha_i$ the sets $\Gamma^\mathrm{s}_{\alpha_i}(x_{i,q})$ and $\Gamma^\mathrm{u}_{\alpha_i}(x_{i,q})$ are defined as
\begin{align}
\Gamma^\mathrm{s}_{\alpha_i}(x_{i,q}) = & \; \alpha_i, \label{eq:setsAlpha1}\\
\Gamma^\mathrm{u}_{\alpha_i}(x_{i,q}) = & \; \mathcal{B}(x_{i,q}),
\label{eq:setsAlpha2}
\end{align}
where $\mathcal{B}(x_{i,q})$ is the set of functional nodes $\beta_e$ connected to $x_{i,q}$. In contrast, for functional nodes $\beta_e$ the sets $\Gamma^\mathrm{s}_{\beta_e}(x_{i,q})$ and $\Gamma^\mathrm{u}_{\beta_e}(x_{i,q})$ are defined as
\begin{align}
\Gamma^\mathrm{s}_{\beta_e}(x_{i,q}) = & \; \mathcal{B}(x_{i,q}),\label{eq:setsBeta1}\\
\Gamma^\mathrm{u}_{\beta_e}(x_{i,q}) = & \; \alpha_i. \label{eq:setsBeta2}
\end{align}
In summary, the factor graph representation described above provides a pictorial representation of the variables and constraints of the CSP.



%

%

\section{Survey Propagation for Resource Allocation}
\label{sec:SP}
In this section, we introduce our proposed resource allocation approach for large networks. To this aim, we first explain the Survey Propagation method from statistical physics and then discuss the details of our algorithm.

\subsection{Survey Propagation}
The Survey Propagation method proposed in \cite{Mezard2002} is a physics-based approach to solve CSPs. 
In a nutshell, Survey Propagation is an iterative message-passing algorithm, where the main idea is to propagate statistical information, or surveys, between the vertices of the factor graph representation of the CSP.
Note that Survey Propagation does not run over the wireless network represented by the graph $\mathcal{G}$, but over the factor graph representation of the resource allocation problem associated to it.
Moreover, the surveys are not physically transmitted messages between the base stations, but are an abstraction of messages ``interchanged'' between the nodes in the factor graph, i.e., between variable and functional nodes. 

The surveys are used 
 to identify the state of the variables in clusters that satisfy the CSP, i.e., whether the variables $x_{i,q}$ are fixed to a certain logical value "0" or "1", or if they are in an indifferent "*" state. In other words, if resource $q$ cannot be allocated to base station $i$ ($x_{i,q}=\text{"}0\text{"}$), if resource $q$ must be allocated to base station $i$ ($x_{i,q}=\text{"}1\text{"}$), or if base station $i$ is indifferent to the allocation of $q$ ($x_{i,q}=\text{"}*\text{"}$).
For example, such cases occur when a neighboring base station has been already allocated resource $q$, when the only available resource for base station $i$ is resource $q$, and when base station $i$ has many available resources and none of its neighbors are already using them, respectively.
When the state of a variable $x_{i,q}$  is fixed to "0" or "1", it can be removed from the problem. This is because, for the satisfaction of the CSP, its value has been set. This means, only the logical value of the variables in the indifferent state remains to be identified. By iteratively repeating this procedure, the original problem is reduced. In every iteration a smaller problem is considered until the space of satisfying solutions is formed by a single cluster.

Specifically, the survey transmitted from a functional node $\alpha_i$ or $\beta_e$ to a connected variable node $x_{i,q}$ is a real number $\eta_{\alpha_i\rightarrow x_{i,q}} \in [0,1]$. 
The survey transmitted from a variable node $x_{i,q}$ to a connected functional node $\alpha_i$ or $\beta_e$ is a triplet $\pi_{x_{i,q}\rightarrow \alpha_i}=(\pi_{x_{i,q}\rightarrow \alpha_i}^\mathrm{u}, \pi_{x_{i,q}\rightarrow \alpha_i}^\mathrm{s}, \pi_{x_{i,q}\rightarrow \alpha_i}^\mathrm{*})$ formed by three real numbers $\pi_{x_{i,q}\rightarrow \alpha_i}^\mathrm{u}$, $\pi_{x_{i,q}\rightarrow \alpha_i}^\mathrm{s}$ and $\pi_{x_{i,q}\rightarrow \alpha_i}^\mathrm{*}$\footnote{In case the considered functional node is $\beta_e$ instead of $\alpha_i$, the subindex is changed accordingly}. 
These surveys can be interpreted as probabilities of warning \cite{Braunstein2005}. Specifically, $\eta_{\alpha_i\rightarrow x_{i,q}}$ can be seen as the probability that functional node $\alpha_i$ warns the variable node $x_{i,q}$ to take the correct value in order to satisfy clause $\alpha_i$.
Similarly, $\pi_{x_{i,q}\rightarrow \alpha_i}^\mathrm{u}$, $\pi_{x_{i,q}\rightarrow \alpha_i}^\mathrm{s}$ and $\pi_{x_{i,q}\rightarrow \alpha_i}^\mathrm{*}$ can be interpreted as the probabilities that variable node $x_{i,q}$ sends a warning to functional node $\alpha_i$ informing it that $x_{i,q}$ cannot satisfy the clause ($\pi_{x_{i,q}\rightarrow \alpha_i}^\mathrm{u}$), that it can satisfy it ($\pi_{x_{i,q}\rightarrow \alpha_i}^\mathrm{s}$) or that it is indifferent ($\pi_{x_{i,q}\rightarrow \alpha_i}^\mathrm{*}$).
All these surveys are recursions calculated from the messages received from neighboring nodes in the factor graph of the CSP. 

For the resource allocation problem, the survey from $\alpha_i$ to $x_{i,q}$ is calculated as
\begin{equation}
\eta_{\alpha_i\rightarrow x_{i,q}} = \prod_{ x_{j,q} \in \mathcal{X}(\alpha_i)\setminus x_{i,q}} {\left[\frac{\pi^\mathrm{u}_{x_{j,q}\rightarrow \alpha_i}}{(\pi_{x_{j,q}\rightarrow \alpha_i}^\mathrm{u} + \pi_{x_{j,q}\rightarrow \alpha_i}^\mathrm{s}+ \pi_{x_{j,q}\rightarrow \alpha_i}^\mathrm{*})}\right]}
\label{eq:eta}
\end{equation}
 \cite{Braunstein2005}. The survey transmitted from $x_{i,q}$ to $\alpha_i$ is calculated as 
\begin{align}
\pi_{x_{i,q}\rightarrow \alpha_i}^\mathrm{u} =& \left[1 - \prod_{\alpha_j \in \Gamma_{\alpha_i}^\mathrm{u}(x_{i,q})}\left(1-\eta_{\alpha_j \rightarrow x_{i,q}}\right) \right] \prod_{\alpha_j \in \Gamma_{\alpha_i}^\mathrm{s}(x_{i,q})}\left(1-\eta_{\alpha_j \rightarrow x_{i,q}}\right), \label{eq:piu} \\
\pi_{x_{i,q}\rightarrow \alpha_i}^\mathrm{s} =& \left[1 - \prod_{\alpha_j \in \Gamma_{\alpha_i}^\mathrm{s}(x_{i,q})}\left(1-\eta_{\alpha_j \rightarrow x_{i,q}}\right) \right] \prod_{\alpha_j \in \Gamma_{\alpha_i}^\mathrm{u}(x_{i,q})}\left(1-\eta_{\alpha_j \rightarrow x_{i,q}}\right), \label{eq:pis}\\
\pi_{x_{j,q}\rightarrow \alpha_i}^\mathrm{*} =& \prod_{\alpha_j \in \Gamma_{\alpha_i}^\mathrm{u}(x_{i,q})}\left(1-\eta_{\alpha_j \rightarrow x_{i,q}}\right) \prod_{\alpha_j \in \Gamma_{\alpha_i}^\mathrm{s}(x_{i,q})}\left(1-\eta_{\alpha_j \rightarrow x_{i,q}}\right) \label{eq:pi*}
\end{align}
\cite{Braunstein2005}, where the sets $\Gamma_{\alpha_i}^\mathrm{u}(x_{i,q})$ and $\Gamma_{\alpha_i}^\mathrm{u}(x_{i,q})$ are defined in \eqref{eq:setsAlpha1} and \eqref{eq:setsAlpha2} for the $\alpha_i \in \mathcal{A}$ functional nodes, and in \eqref{eq:setsBeta1} and \eqref{eq:setsBeta2} for the $\beta_e \in \mathcal{B}$ functional nodes.

The surveys are calculated for each node in the factor graph according to the surveys received from neighboring nodes. That means, their calculation is an iterative process that concludes when the surveys have reached a steady value, i.e., the difference between the surveys calculated in two consecutive iterations is less than a given threshold $\epsilon$ for all nodes, or when the maximum number  $T_\mathrm{SP}$ of iterations has been reached. 
Algorithm \ref{alg:SP} summarizes the procedure used to calculate the surveys.  
First, the surveys are randomly initialized for all the nodes in the factor graph, and the maximum number of iterations $T_\mathrm{SP}$, the convergence threshold $\epsilon$, and the indicator variable convergenceSP are set (lines 1-2).
While the surveys have not yet converged, a random permutation is generated for the order in which an individual survey will be updated (lines 3-4). Using this order, the surveys are updated using Eq. \eqref{eq:eta}-\eqref{eq:pi*} (line 5). Then, for every survey, we evaluate whether the value has converged according to the predefined threshold $\epsilon$ (lines 6-12). 
If the condition is not fulfilled for all the nodes in the factor graph, the procedure is repeated (line 10). On the contrary, if the values of $\eta_{\alpha_i\rightarrow x_{i,q}}$ have reached convergence, then these values are used as surveys (line 15).
If the values of the surveys do not converge and the maximum number of iterations $T_\mathrm{SP}$ is reached, the surveys are initialized once again and Algorithm \ref{alg:SP} runs once more.
The convergence properties of Survey Propagation are discussed in Section \ref{sec:treeLike}.

\begin{algorithm}[t]
\DontPrintSemicolon
\scriptsize
	Randomly initialize the surveys $\eta_{\alpha_i\rightarrow x_{i,q}}$ for all $\alpha_i, \beta_e \in \Gamma$ and $x_{i,q} \notin \mathcal{X}^* $\;
	Initialize $T_\mathrm{SP}, \; \epsilon$ \;
	Set $\text{convergenceSP = 0}$ and counter $t_\mathrm{SP}=1$\;
	 \While {$(t_\mathrm{SP} \leq T_\mathrm{SP}) \;\mathbf{and} \; (\text{convergenceSP = 0})$}	{
		Generate a random permutation for the order in which the surveys $\eta_{\alpha_i\rightarrow x_{i,q}}$ will be updated \;
		Update the surveys $\eta_{\alpha_i\rightarrow x_{i,q}}$ using the previously defined order \tcp*{Eq. \eqref{eq:eta}-\eqref{eq:pi*}}
		Set $\text{convergenceSP = 1}$ \;
		\For {every survey $\eta_{\alpha_i\rightarrow x_{i,q}}$}{
			\If {$ |\eta_{\alpha_i\rightarrow x_{i,q}}(t_\mathrm{SP})-\eta_{\alpha_i\rightarrow x_{i,q}}(t_\mathrm{SP}-1)| > \epsilon$}{
				Set $\text{convergenceSP = 0}$ \;
				Go to line \ref{alg:SPrepeat} \;
			}
		}
		Set  $t_\mathrm{SP} =  t_\mathrm{SP} +1$ \label{alg:SPrepeat} \;
	}
	\textbf{return} $\eta_{\alpha_i\rightarrow x_{i,q}}$ \;
\caption{Surveys update} \label{alg:SP}
\end{algorithm}


\subsection{Resource Allocation Algortihm}
\label{sec:algo}
In this section, we describe the proposed algorithm to find the resource allocation solution that minimizes the interference in large wireless network.
The algorithm is based on the Survey Propagation method of \cite{Mezard2002} and the corresponding decimation algorithm described in \cite{Braunstein2005}.
It is an iterative approach that aims at fixing the value of the variables $x_{i,q}$ based on the exchanged survey messages. In other words, in each iteration $t=1,...,T$, where $T$ is the maximum number of iterations, the algorithm fixes the allocation of a resource $q$ to one or more base stations $i$ according to the values of the surveys. 

\begin{algorithm}[t]
\scriptsize
From the given wireless network, build the corresponding CSP and factor graph \tcp*{Sec. \ref{sec:probFormulation}}
Initialize parameters $T$, $T'$\;
Set $t=1$, $t'=1$ \;
\While {$ \mathcal{X}^*\neq \mathcal{X}$}{
	\While {$(t \leq T) \land (\mathcal{X}^*\neq \mathcal{X}) \land (t' \leq T') $}{
		Calculate the surveys $\eta_{\alpha_i\rightarrow x_{i,q}}$  \tcp*{Alg. \ref{alg:SP}}
		\eIf {the surveys converge to a steady value}{
			\eIf {all the surveys have values larger than zero}{
				Set $t'=1$	\;
				 Calculate the biases $W^+_{x_{i,q}}$, $W^-_{x_{i,q}}$ of each variable node $x_{i,q}$ \tcp*{ Eq. \eqref{eq:bias1}-\eqref{eq:bias5}}
				 Find variable node $x^*_{i,q}$ with largest bias difference $|W^+_{x_{i,q}}-W^-_{x_{i,q}}|$\;
			
				Set $x^*_{i,q}=1$ and $x_{i,r}=0,\; \forall r=1,...,Q, r\neq q$\;
				 Set $x_{j,q}=0,\; \forall j, (i,j) \in \mathcal{E}$ \label{updateGraph}\;
				Remove variable node $x^*_{i,q}$ and its associated edges $e=(i,j),\; \forall j\in \mathcal{V}$ from the factor graph\;
				 Find subset $\mathcal{V}^1$ of base stations that have only one option left\;
				\While {$\mathcal{V}^1\neq \emptyset$}{
					\For {every $i\in \mathcal{V}^1$}{
						 Allocate the available resource $q$ to base station $i$, $x_{i,q}=1$ \;
						 Set $x_{j,q}=0,\; \forall j, (i,j) \in \mathcal{E}$ \;
						 Remove variable node $x_{i,q}$ and its associated edges $e=(i,j),\; \forall j\in \mathcal{V}$ from the factor graph\;
					}
					Update subset $\mathcal{V}^1$\;
				}
				Update $\mathcal{X}^+$ and set $t=t+1$	\;
			}{
				Use a greedy heuristic for the allocation of resources to the remaining nodes, e.g., \cite{Ramanathan1999}\;
			}
		}{
			Set $t'=t'+1$\;
		}
	}
	\If {$t'=T'$}{
		Set $t'=1$\;
		Randomly select one base station $i$ from the set of base stations with the largest number of neighbors\;
		 Assign one of the available resources $q$ and set $x^*_{i,q}=1$ and $x_{i,r}=0,\; \forall r=1,...,Q, r\neq q$	\;
		 \textbf{go to} line \ref{updateGraph}\;
	}
}
\caption{Proposed resource allocation algorithm } \label{alg:main} 
\end{algorithm}

The proposed algorithm is presented in Algorithm \ref{alg:main}.
The first step is, for the given network, to transform the resource allocation problem into a CSP and build the corresponding factor graph, as explained in Sec. \ref{sec:probFormulation} (line 1).
Second, the algorithm parameters and variables are initialized (lines 2-3).
Let $\mathcal{X}^*\subseteq \mathcal{X}$ be the set of variable nodes $x_{i,q}$ whose value has been fixed.
The execution continues while there are still base stations without an allocated resource $q$, i.e., $\mathcal{X}^*\neq \mathcal{X}$, while the number of iterations is less than the maximum value $T$,  and while the number of times the Survey Propagation method has been run in a single iteration does not exceed a maximum value $T'$ (line 4-5).
The last condition comes from the fact that, as mentioned in Sec. \ref{sec:SP} and \ref{sec:treeLike}, the convergence of the Survey Propagation method is not guaranteed. Therefore, we include the variable $T'$ as the maximum number of times the Survey Propagation method can be run in a single iteration.
Once the surveys $\eta_{\alpha_i\rightarrow x_{i,q}}$ converge to a steady value (line 6-7), we evaluate if their value is larger than zero (line 8). If this is not the case, then we face the trivial solution and the greedy heuristic in \cite{Ramanathan1999} is used to allocate the resources to the remaining base stations $i$ (line 26-27). 
If the surveys $\eta_{\alpha_i\rightarrow x_{i,q}}>0$ then the bias of each variable $x_{i,q}$ towards the possible logical states "0"  and "1" is calculated (line 10). The biases represent, according to the surveys, how sure is base station $i$ about being allocated (or not) resource $q$.
The bias towards the logical state "0" is denoted as $W^-_{x_{i,q}}$ and the bias towards "1" is denoted by $W^+_{x_{i,q}}$. Both quantities are calculated as 
\begin{align}
W^-_{x_{i,q}}= & \frac{\pi^-_{x_{i,q}}}{\pi^+_{x_{i,q}}+\pi^-_{x_{i,q}}+\pi^0_{x_{i,q}}}, \label{eq:bias1}\\
W^+_{x_{i,q}}= & \frac{\pi^+_{x_{i,q}}}{\pi^+_{x_{i,q}}+\pi^-_{x_{i,q}}+\pi^0_{x_{i,q}}} \label{eq:bias2}
\end{align}
\cite{Braunstein2005}, where $\pi^+_{x_{i,q}} $, $\pi^-_{x_{i,q}} $ and $\pi^0_{x_{i,q}} $ are given by 
\begin{align}
\pi^+_{x_{i,q}} = &\left[1 - \prod_{\alpha_i \in (\Gamma(x_{i,q})\cup \mathcal{A})} (1-\eta_{\alpha_i\rightarrow x_{i,q}})\right]\prod_{\beta_e \in (\Gamma(x_{i,q})\cup \mathcal{B})} (1-\eta_{\beta_e\rightarrow x_{i,q}}) \label{eq:bias3} \\
\pi^-_{x_{i,q}}  = &\left[1 - \prod_{\beta_e \in (\Gamma(x_{i,q})\cup \mathcal{B})} (1-\eta_{\beta_e\rightarrow x_{i,q}})\right]\prod_{\alpha_i \in (\Gamma(x_{i,q})\cup \mathcal{A})} (1-\eta_{\alpha_i\rightarrow x_{i,q}}) \label{eq:bias4} \\
\pi^0_{x_{i,q}} = & \prod_{\alpha_i \in \Gamma(x_{i,q})} (1-\eta_{\alpha_i\rightarrow x_{i,q}}). \label{eq:bias5}
%
\end{align}
The next step is then to find the variable node $x^*_{i,q}$ with the largest absolute difference $|W^+_{x_{i,q}}-W^-_{x_{i,q}}|$ between biases (line 11). 
Resource $q$ is allocated to base station $i$ by setting $x^*_{i,q}=1$. 
Furthermore, the use of resource $q$ is marked as forbidden for the neighboring base stations $j$ for $(i,j)\in \mathcal{E}$ (lines 12-13)
After fixing the value of $x^*_{i,q}$, the variable node and its edges are removed from the factor graph (line 14).
As the algorithm iterates and variables are fixed, the set of resources that can be allocated to other base stations is also reduced. This is because the allocation of resource $q$ to base station $i$ forbids the use of the same resource in any neighboring base station $j$.
In particular, we are interested in the subset $\mathcal{V}^\mathrm{1}\subseteq \mathcal{V}$ of base stations that have only one option left, i.e., there is only one resource which can be allocated to them in order to avoid the reuse of resources (line 15).
For these base stations, the available resource $q$ is allocated and its use on neighboring base stations is forbidden. Additionally, the corresponding variables nodes $x_{i,q}$ are removed from the factor graph (lines 16-23).
Next, the set $\mathcal{X}^*$ is updated (line 24). If all the variables nodes in the factor graph have been set, then the algorithm returns this assignment as the resource allocation solution. Otherwise, the counter $t$ is increased and another iteration starts. 
If after $T'$ consecutive runs the surveys do not converge to a steady value, we randomly select one base station $i$ from the set of base stations with the largest number of neighbors. For this base station $i$ we allocate one of the available resources, i.e, $x_{i,q}=1$ and update the factor graph (lines 33-36).

\section{Convergence of Survey Propagation}
\label{sec:conv}
\subsection{Convergence }
\label{sec:treeLike}
As reported by the proposers of Survey Propagation, there is no general proof of convergence of the algorithm for arbitrary CSPs \cite{Mezard2002}. 
This is because the survey updates presented in the previous Sec. \ref{sec:SP} are based on the cavity method from Statistical Physics. The cavity method is a state-of-the-art non-rigorous approach to calculate minimum energy states in spin glasses. Turning the cavity method into a rigorous theory is an open research question \cite{Braunstein2006, Braunstein2005}.
As in any message-passing algorithm, the assumption that the messages transmitted among neighboring nodes in the factor graph representation are statistically independent, can only be guaranteed when the considered graphs have a tree structure. 
This means, the solution of the cavity method, and in turn, the Survey Propagation approach is exact only for factor graphs with a tree structure.
Nevertheless, numerical results have shown that the application of Survey Propagation is not limited to only trees, but is able to find solutions, and outperform other state-of-the-art techniques, in cases when the considered graph has a local tree-like structure \cite{Mezard2002, Braunstein2005}. 
The intuition behind this is that when the considered network has a tree-like structure, the length of the existing loops in the graph grows with the number of considered nodes. Thus for large networks the correlation between the exchanged surveys among neighboring nodes is reduced.

\subsection{Tree-Like Structure in Wireless Networks}
As discussed in the previous Sec. \ref{sec:treeLike}, numerical results have shown that Survey Propagation is able to find a solution to CSP when the corresponding factor graph has a tree-like structure.
In this section, we present the Gromov's $\delta$-hyperbolicity of a graph, which can be used to measure the "tree-likeness" of the considered wireless network.

The Gromov's $\delta$-hyperbolicity aims at characterizing the tree-likeness of a metric space in terms of its distance \cite{Wei2012, Adcok2013}. Trees have a hyperbolicity $\delta=0$. Therefore, the smaller the $\delta$-hyperbolicity of a graph, the more tree-like it is.
For a graph $G=(\mathcal{V},\mathcal{E})$, the value of $\delta$ ranges from 0 to half of the graph diameter. 
Similar to \cite{Wei2012, Adcok2013}, we consider the Gromov 4-point $\delta$-hyperbolicity as the metric of the tree-likeness of graph $G$. 

To define the $\delta$-hyperbolicity, consider that the graph $G$ can be viewed as metric space $(\mathcal{V}, d_G)$, where $\mathcal{V}$ is the set of vertices and $d_G$ is the geodesic distance between two vertices $i$ and $j$. Then, $\delta$-hyperbolicity is defined as  in \cite{Adcok2013}
\begin{theorem}
Let $0\leq \delta < \infty$. $(\mathcal{V}, d_G)$ is 4-point $\delta$-hyperbolic if and only if for any four vertices $i,j,k,l \in \mathcal{V}$ ordered such that $d_G(i,j) + d_G(k,l) \geq d_G(i,k) + d_G(j,l) \geq d_G(i,l) + d_G(j,k)$, the following inequality holds
\begin{equation} 
(d_G(i,j) + d_G(k,l) ) - (d_G(i,k) + d_G(j,j)) \leq 2\delta.
\end{equation}
\end{theorem}
The $\delta$ of $G$ is then the minimum value of $\delta$ such that $G$ is $\delta$-hyperbolic.

\section{Numerical Evaluation}
\label{sec:simResults}

In this section, we evaluate the performance of the proposed resource allocation algorithm via numerical simulations.
For the evaluation, 500 independent realizations are considered. Each realization corresponds to the generation of a random wireless network. We consider two models for the generation of the networks, namely, an Erd\"os-Renyi model and a random geometric one. 
The Erd\"os-Renyi model provides us with a theoretical test-bed as it is the model commonly used in statistical physics, while the random geometric model is used to generate more realistic wireless communication networks.
To ensure that the CSPs lie in the solvable interval $\theta_d \leq \theta \leq \theta_c$ and based on the results in  \cite{Zdeborova2007}, we assume that the probability that two base stations interfere with each other in the Erd\"os-Renyi model is $4.5/I$, where $I$ is the number of considered base stations.
For the random geometric model we assume the $I$ base stations are distributed over an area of one square kilometer following a Poisson Point process and the neighborhoods are defined according to the channel conditions.
%
%
The channel gains $|h_{i,j}|$ describing the channel between neighboring nodes $i$ and $j$ are taken from an independent and identically distributed Rayleigh fading process with zero mean, unit variance and a path loss exponent of three. 
Additionally, zero mean additive white Gaussian noise with variance $\sigma^2=1$ is considered.
For data transmission, the nodes set the transmit power to $p^\mathrm{Tx}=100\mathrm{mW}$.
Two nodes $i$ and $j$ are said to be neighbors if the received power $p^\mathrm{Rx}_{i,j}$ of the interference signal from $i$ to $j$ is larger than $\mu=-75\mathrm{dBm}$.
For the communication, a bandwidth $B=20$MHz and a time slot duration of 10ms is considered. These time-frequency resources are divided into $Q$ orthogonal resource pools. 

For Survey Propagation, we consider a threshold $\epsilon=10^{-3}$ and a maximum of $T_\mathrm{SP}=10$ iterations for the convergence of the surveys. Additionally, for our proposed algorithm the maximum number $T$ of iterations is set to $T=IQ$ iterations.
To evaluate the performance of our proposed method, we compare it to the following approaches:
\begin{itemize}
\item Belief Propagation \cite{Braunstein2005}: In this case, the resource allocation problem is solved using the Belief Propagation algorithm described in \cite{Braunstein2005}. Belief Propagation is a message passing algorithm in which only the probability of each variable node $x_{i,q}$ taking the logical "0" or the logical "1" value is calculated. 
\item Maximum Neighbor First (MNF) \cite{Ramanathan1999}: A greedy heuristic for resource allocation in which the nodes with the larger number of neighbors are served first. This is, the resource pools are iteratively allocated to one node at a time and the nodes with the largest number of neighbors are considered first. It is called a greedy allocation because for each node $i$, the resource $q$ with the smallest index is assigned without violating any of the constraints in the CSP.
\item Progressive Maximum Neighbor First (PMNF) \cite{Ramanathan1999}: A greedy heuristic like MNF. The only difference is that after allocating a resource $q$ to a node $i$, the node is removed from the factor graph. As a consequence, the number of neighbors changes after each allocation.
\item Greedy with Random Ordering: In this approach a greedy allocation is also considered. However, in each iteration, a node $i$ is randomly selected for the allocation of resource $q$ regardless of the number of neighbors it has. As in the previous approaches, i.e., MNF and PMNF, the resource $q$ with the smallest index which does not violate any of the constraints in the CSP, is assigned to node $i$.
\end{itemize}

\begin{figure}
\centering
\begin{subfigure}{0.46\textwidth}
	\resizebox{1\textwidth}{!}{\input{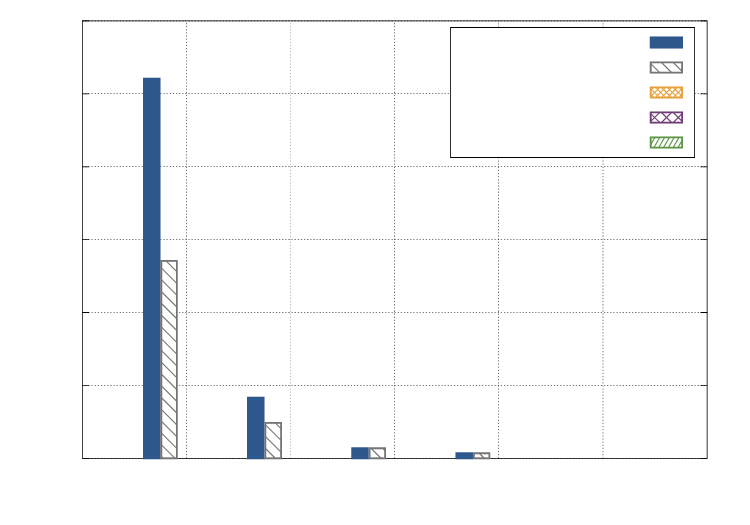}}
    \caption{Percentage of zero interference allocations.}
    \label{fig:convergence_ER_N100-500_Q3}
\end{subfigure}
\hfill
\begin{subfigure}{0.46\textwidth}
\resizebox{1\textwidth}{!}{\input{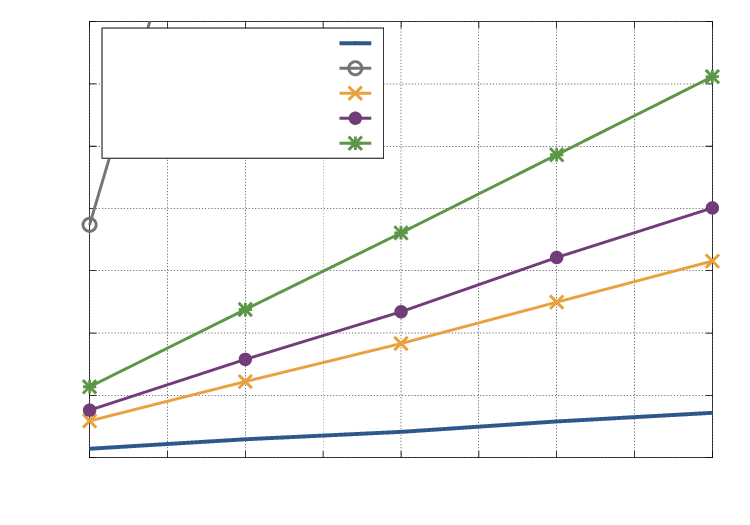}}
    \caption{Average number of interference links.}
    \label{fig:nInterference_ER_N100-500_Q3}
\end{subfigure}
\hfill
\begin{subfigure}{0.46\textwidth}
\resizebox{1\textwidth}{!}{\input{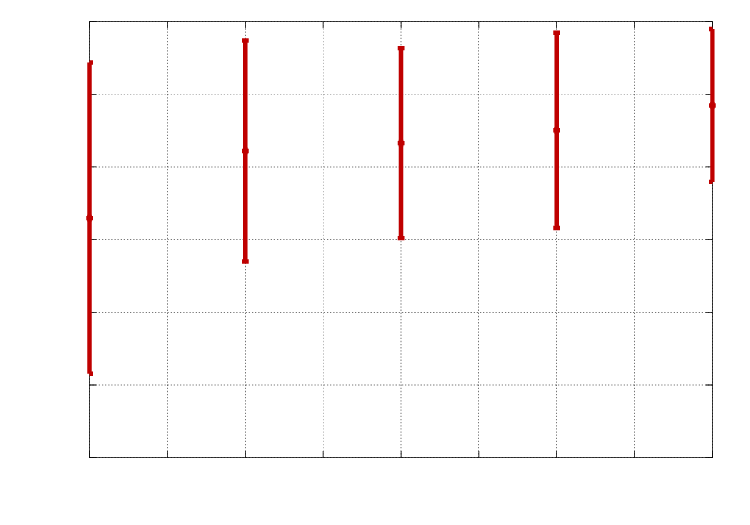}}
    \caption{Average number of neighboring base stations.}
    \label{fig:conect_ER_N100-500_Q3}
\end{subfigure}
\caption{Results for a network generated using the Erd\"os-Reny model with a varying number $I$ of base stations and $Q=3$ resources.}
\label{fig:ER_N100-500_Q3}
\end{figure}

In Fig. \ref{fig:ER_N100-500_Q3}, a scenario with $Q=3$ resources is considered. The networks are randomly generated using the Erd\"os-Renyi model and we evaluate the performance of the proposed algorithm for different network sizes.
Fig. \ref{fig:convergence_ER_N100-500_Q3} shows how many times, in percentage, a zero interference allocation solution is  found. A zero interference allocation solution means that no neighboring base stations use the same resource, and therefore, do not interfere with each other, i.e., $C=0$. To compare the performance of the different schemes, we vary the average number $I$ of base stations in the network from $I=100$ to $I=500$. 
It can be seen that our proposed approach outperforms the reference schemes in all the network size range. Moreover, the greedy-based heuristics, i.e., MNF, PMNF and random order, are not able to find zero interference allocations.
Note that the percentage of zero interference allocation decreases for larger network sizes. 
This is due to the fact that a higher number of base stations increases the complexity of the resource allocation problem. 
The solvabillity of the resource allocation problem depends on the particular topology of the generated network and the number  $Q$ of available resources.
The larger the network size, the larger the average number of neighbors a base station has. 
Fig. \ref{fig:conect_ER_N100-500_Q3} shows both, the average number of neighbors and their standard deviation.
Physicists have identified that for Erd\"os-Renyi networks with $Q=3$ resources the solution space is clusterized (see Section \ref{sec:CSP}) when the average number of neighbors a base station has is larger than or equal to four. Furthermore, when the average number of neighbors is larger than 4.69 the problem is not longer solvable \cite{Zdeborova2007}. This means, the considered scenario is already in the ``hard'' region where simple heuristics fail to find zero interference allocations.
The good performance of the proposed approach is based on the fact that the allocation of resources is not done by considering only the number of neighbors each base station has, like the MNF and PMNF approaches do. The exchange of surveys in the proposed approach allows us to consider the impact an allocation will have on the satisfiability of the whole problem. Although Belief propagation is a message-passing algorithm, like our proposed approach, its performance is lower since it assumes a tree structure for the network which, in general, does not hold.

The solvability of a CSP depends on the topology of the network and the available number of resources. Therefore, finding zero interference allocations might not be possible for some real world networks. For this reason, it is of practical relevance to evaluate the average number of interference links. That is, when a zero interference allocation cannot be found, how many pairs of base stations will interfere with each other. Fig. \ref{fig:nInterference_ER_N100-500_Q3} shows the number of interference links for the same scenario of $Q=3$ resources and a varying number of base stations. We can see that for $I=500$ our proposed approach achieves 76\%, 81\% and 87\% less conflicts than MNF, PMNF and the random order approaches, respectively. Note that the belief propagation algorithm, as presented in \cite{Braunstein2005}, stops its execution whenever a pair of base stations have no other option than to use the same resource. For this reason, the number of interference links is approximately ten times larger than the proposed scheme for $I=500$.

\begin{figure}[htp]
\centering
\begin{subfigure}{0.46\textwidth}
	\resizebox{1\textwidth}{!}{\input{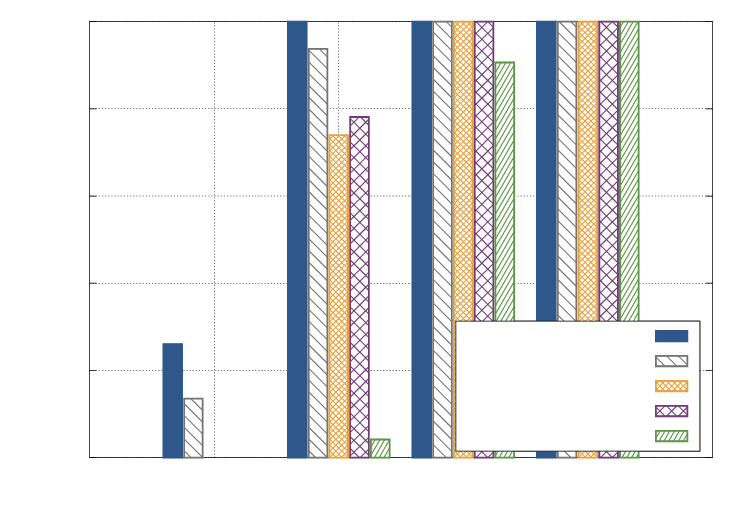}}
    \caption{Percentage of zero interference allocations.}
     \label{fig:Convergence_ER_N100_Q3_Q10}
\end{subfigure}
\hfill
\begin{subfigure}{0.46\textwidth}
\resizebox{1\textwidth}{!}{\input{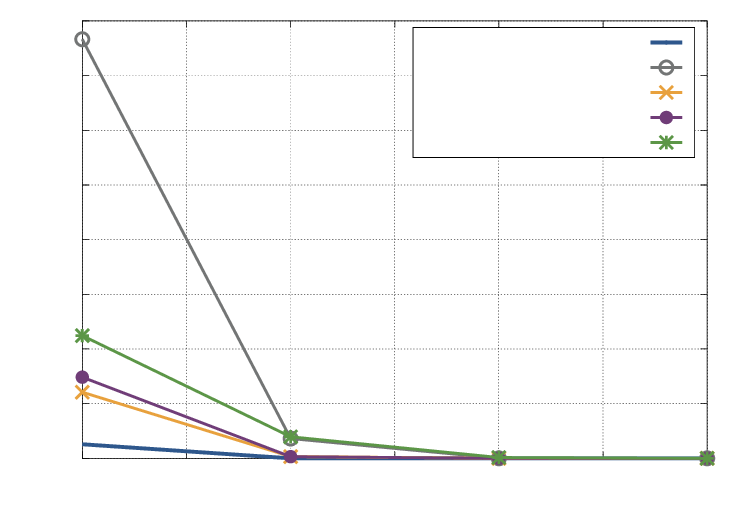}}
    \caption{Average number of interference links.}
     \label{fig:nInterferers_ER_N100_Q3_Q10}
     \end{subfigure}
\caption{Results for a network generated using the Erd\"os-Reny model with $I=100$ base stations and a varying number $Q$ of resources.}
\label{fig:ER_N100_Q3-10}
\end{figure}

In Fig. \ref{fig:ER_N100_Q3-10}, we evaluate the performance of the considered approaches for Erd\"os-Renyi networks when $I=100$ base stations and a variable number $Q$ of resources are considered. 
When $Q$ increases, the performance of all the considered algorithms improves. This is because the difficulty of the allocation problem is reduced when more resources are available, even if at the same time more resources mean a larger dimensionality of the problem.
However, the more scarce the resources, the larger the gains achieved by our proposed approach.
Fig. \ref{fig:Convergence_ER_N100_Q3_Q10} shows the percentage of zero interference allocations as a function of the number of resources. We can see that for $Q=4$, our proposed approach is able to find zero interference allocations for all the considered networks. The belief propagation approach, MNF, PMNF and the random order are able to find them only 93\%, 74\%, 78\%, and 4\% of the times.
In Fig. \ref{fig:nInterferers_ER_N100_Q3_Q10} the number of interfering links is shown. We can see that reducing the number of resources has a dramatic impact on the performance of the reference schemes.
For the proposed approach, the number of interference links is already zero for $Q=4$. The reference schemes are only able to achieve this value when $Q\geq5$.

\begin{figure}[htp]
\centering
\begin{subfigure}{0.46\textwidth}
	\resizebox{1\textwidth}{!}{\input{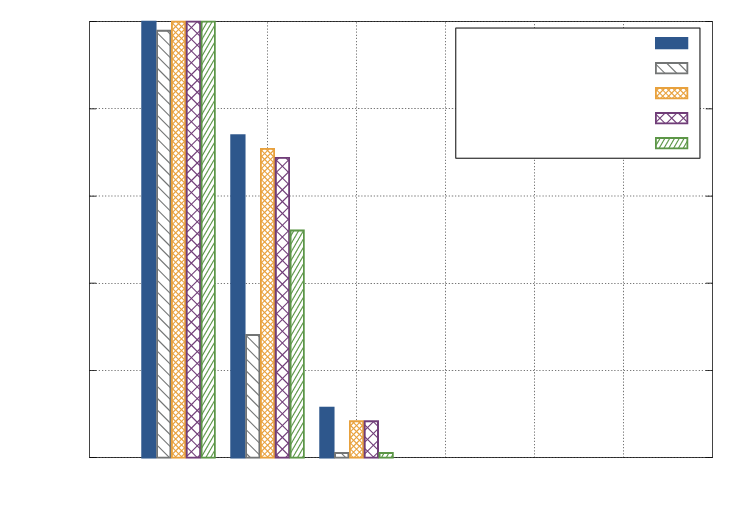}}
    \caption{Percentage of zero interference allocations.}
    \label{fig:convergence_N50-300_Q5}
\end{subfigure}
\hfill
\begin{subfigure}{0.46\textwidth}
\resizebox{1\textwidth}{!}{\input{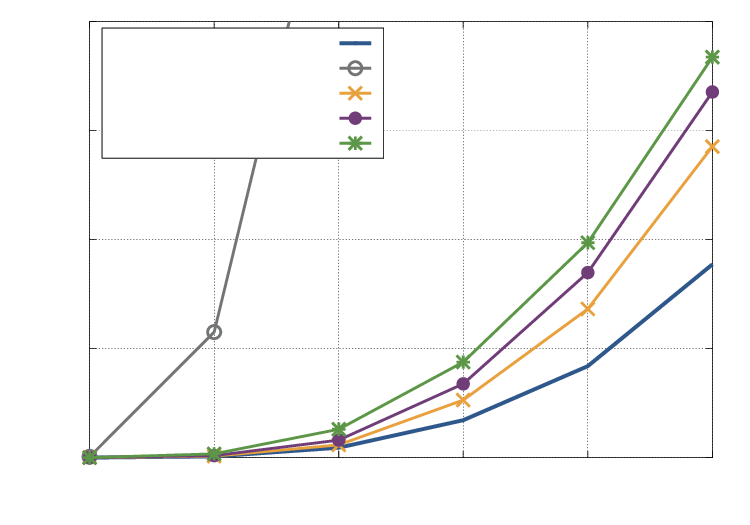}}
    \caption{Average number of interference links.}
    \label{fig:nInterference_N50-300_Q5}
\end{subfigure}
\caption{Results for a geometric random network with a varying number $I$ of base stations, $Q=5$ resources and power threshold $\mu=-75$dBm.}
\label{fig:N50-300_Q5}
\end{figure}


The random geometric model to generate the wireless networks is considered next.
 In Fig. \ref{fig:N50-300_Q5}, the number of base stations is varied from $I=50$ to $I=300$, and $Q=5$ resources are assumed to be available. 
Fig. \ref{fig:convergence_N50-300_Q5} shows the percentage of zero interference allocations versus the network size.
As expected, the percentage of zero interference allocations is reduced when the number of base stations increases. Note that in contrast to Fig. \ref{fig:convergence_ER_N100-500_Q3}, the reference schemes MNF, PMNF and random order  outperform belief propagation. This is because, the tree-likeness of these networks is much smaller compared to the tree-likeness of networks generated using the Erd\"os-Renyi model, i.e., smaller $\delta$-hyperbolicity. Nevertheless, our proposed scheme outperforms all the references approaches. 
The number of interference links is presented in Fig. \ref{fig:nInterference_N50-300_Q5}. It can be seen that the proposed approach has consistently the lowest number of interference links. Specifically, for $I=300$ it achieves 95\%, 38\%, 47\%, and 52\% less conflicts than belief propagation, MNF, PMNF and random order, respectively. 

  \begin{figure}[H]
  \centering
  	\resizebox{0.46\columnwidth}{!}{\input{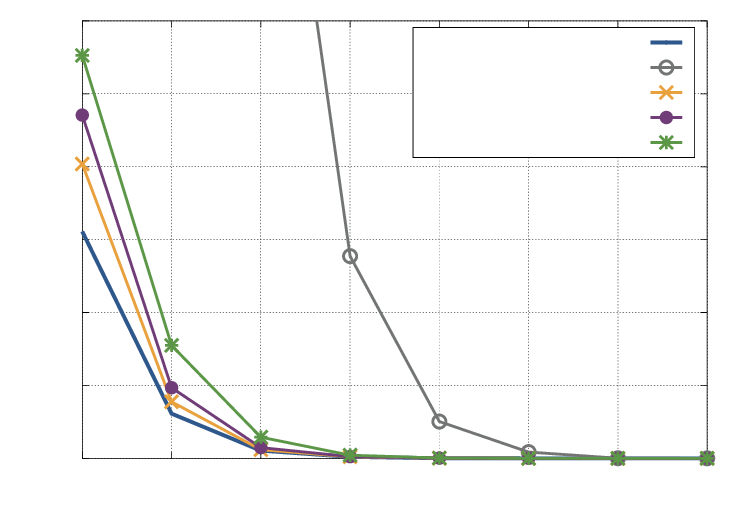}}
  \caption{Average number of interference links in a geometric random network with $I=100$ base stations and power threshold $\mu=-75$dBm.}
  \label{fig:Interferers_N100_Q3_Q10}
  \end{figure} 
  
The impact of the number of resources on the average number of interference links is presented in Fig. \ref{fig:Interferers_N100_Q3_Q10}. As in Fig. \ref{fig:N50-300_Q5} the networks are generated using the geometric random model with a power threshold $\mu=-75$dBm.
From the average number of interference links and the results in Fig. \ref{fig:N50-300_Q5}, it is clear that $Q=3$ resources are not enough to obtain zero interference allocations. Nevertheless, our proposed approach achieves 23\% less conflicts than the MNF reference scheme. This gain is expected to increase for larger networks.
As for the case of Erd\"os-Renyi networks, the highest gains of our proposed approach are achieved when the problem becomes more difficult, i.e., when the number of resources is reduced.


  \begin{figure}[t]
\centering
\begin{subfigure}{0.46\textwidth}
	\resizebox{1\textwidth}{!}{\input{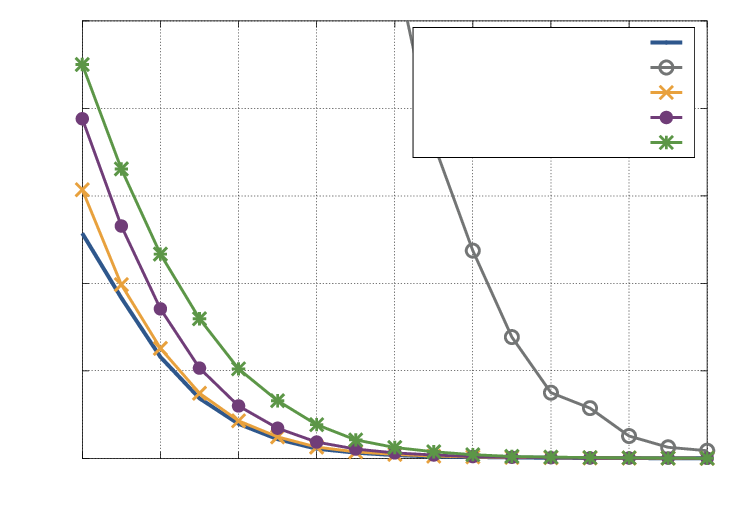}}
    \caption{Average number of interference links.}
    \label{fig:Interferers_N100_Q6_Thr-116_-100}
\end{subfigure}
\hfill
\begin{subfigure}{0.46\textwidth}
\resizebox{1\textwidth}{!}{\input{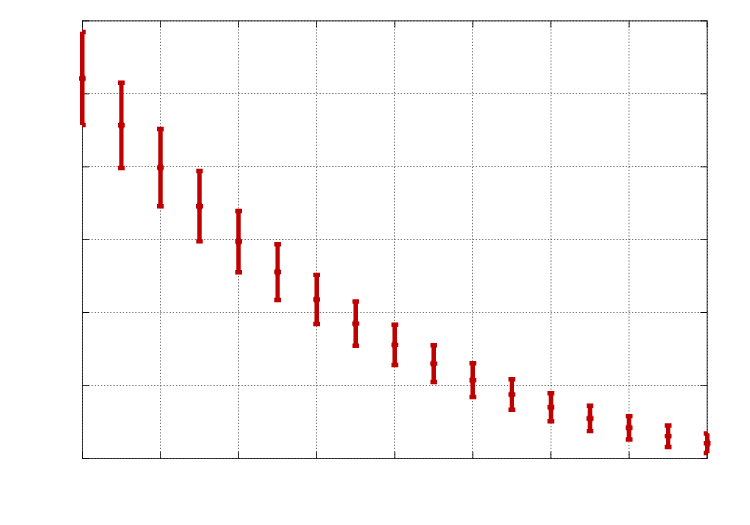}}
    \caption{Average number of neighboring base stations.}
    \label{fig:conect_Thr-116-100_Q6}
\end{subfigure}
\caption{Results for a geometric random network with $I=100$ base stations, $Q=6$ resources and a varying power threshold $\mu$.}
\label{fig:N100_Q6_Thr-116_-100}
\end{figure}

Finally, in Fig. \ref{fig:N100_Q6_Thr-116_-100} we evaluate the impact of the power threshold $\mu$. The smaller $\mu$, the larger the average number of neighbors per base station, as shown in Fig. \ref{fig:conect_Thr-116-100_Q6}. 
The power threshold $\mu$ impacts the topology of the network and reduces its tree-likeness. 
As previously discussed, a larger number of average neighbors increases the complexity of the allocation problems.
For this reason, we have extended in this case the number of available resources to $Q=6$.  
It can be seen that for $\mu=-86$dBm and $\mu=-80$dBm, our proposed approach achieves 16\%, and 14\% less conflicts than the reference scheme MNF, respectively.
From the simulation results presented in this section, it is clear that although survey propagation has been mainly used by physicists to solve networks generated using the Erd\"os-Renyi model, it is also able to outperform reference schemes when more realistic wireless networks models are considered.
These results show that our proposed approach is a promising tool for the efficient allocation of communication resources in wireless communications networks. For a fixed number of resources, it is able to serve a larger number of nodes in the network without introducing more interference to the system.
\section{Conclusions}
\label{sec:conclusions}
Motivated by the scalability challenges posed by the next generation of wireless networks, in this paper the resource allocation problem for large wireless networks is studied. To address this problem, a novel approach based on statistical physics is investigated. 
Specifically, we propose a model of the wireless network inspired by spin glasses, a type of disordered physical systems.
Based on this model, we show how the resource allocation problem can be written as a constraint satisfiability problem.
This formulation allows us to exploit the survey propagation method from statistical physics to find resource allocation solutions that minimize the interference.
The main advantage of our proposed approach is that it can be applied to networks with a large number of communication nodes. 
Although physicists have mainly applied survey propagation to networks generated using the Erd\"os-Renyi model, our work shows that this approach also outperforms reference schemes when more realistic models for wireless communication networks are considered. 
For a fixed number of resources, our approach is able to serve a larger number of nodes in the network without introducing more interference to the system.
In our opinion, these results should be the starting point and, at the same time, motivate a joint effort of both communities. From this joint work, an improvement of the available methods from statistical physics can be achieved and a new set of tools to meet the challenges posed by next generation wireless communication networks can be developed. Future work aims at the consideration of joint resource and power allocation problems a well as dynamic environments.




\bibliographystyle{IEEEtran}


\footnotesize
\bibliography{biblio}


\end{document}